\documentclass[aps,floats,twocolumn]{revtex4}
\usepackage{tabularx,graphicx}
\usepackage{epsfig}
\usepackage{amsmath}
\usepackage{bbm}
\usepackage{graphicx}

\begin{document}

\title{Screening effects on the excitonic instability in graphene}
\author{J. Gonz\'{a}lez \\}
\address{Instituto de Estructura de la Materia,
        Consejo Superior de Investigaciones Cient\'{\i}ficas, Serrano 123,
        28006 Madrid, Spain}

\date{\today}

\begin{abstract}
We investigate the excitonic instability in the theory of Dirac fermions 
in graphene with long-range Coulomb interaction. We analyze the electron-hole 
vertex relevant for exciton condensation in the ladder approximation, showing 
that it blows up at a critical value of the interaction strength 
$\alpha = e^2/4\pi v_F$ sensitive to further many-body 
corrections. Under static screening of the interaction, we find that taking 
into account electron self-energy corrections increases the critical coupling 
to $\alpha_c \approx 2.09$, for a number $N = 4$ of two-component Dirac 
fermions. We show that the dynamical screening of the interaction has 
however the opposite effect of enhancing the instability, which turns out to 
develop then at $\alpha_c \approx 0.99$ for $N = 4$, bringing the question of
whether that critical value can be reached by the effective coupling in real 
graphene samples at the low-energy scales of the exciton condensation.

\end{abstract}

\maketitle



{\em Introduction.---} 
The discovery of graphene, the material made of a one-atom-thick carbon layer, 
has attracted a lot attention as it provides the realization of a system
where the electrons have conical valence and conduction bands, therefore
behaving at low energies as massless Dirac fermions\cite{geim,kim,rmp}. 
This offers the possibility of employing the new material as a test ground 
of fundamental concepts in theoretical physics, since the interacting electron
system represents a variant of strongly coupled quantum electrodynamics (QED)
affording quite unusual effects\cite{nil,fog,shy,ter}.

A remarkable feature of such a theory is that a sufficiently strong Coulomb 
interaction may open a gap in the electronic spectrum. 
This effect, already known from the study of QED \cite{appel}, was first 
addressed in graphene in the context of the theory with a large number $N$ of 
fermion flavors\cite{khves,gus,ale,son}. The existence of a critical point 
for exciton condensation was also suggested from second-order calculations 
of electron self-energy corrections\cite{vafek}.
More recently, Monte Carlo 
simulations of the field theory have been carried out on the graphene 
lattice\cite{drut1,hands}, showing 
that the chiral symmetry of the massless theory can be broken at the 
physical value $N = 4$ above a critical interaction strength
$\alpha_c \approx 1.08$ \cite{drut1}. The possibility of
exciton condensation has been also studied in the ladder 
approximation\cite{gama,fer,me,brey}, leading in the case of static screening of 
the interaction to an estimate of the critical coupling 
$\alpha_c \approx 1.62$ for $N = 4$ \cite{gama}. Lately, the resolution 
of the Schwinger-Dyson formulation of the gap equation has revealed that
the effect of the dynamical polarization can 
significantly lower the critical coupling for exciton condensation, 
down to a value $\alpha_c \approx 0.92$ for $N = 4$\cite{ggg}.

In this paper we take advantage of the renormalization properties of the 
Dirac theory in order to assess the effect of different many-body 
corrections to the excitonic instability. In this respect, the renormalization 
of the quasiparticle properties can have a significant impact, mainly through 
the increase of the Fermi velocity at low energies\cite{khves2,sabio}. 
Thus, we will consider the renormalization of the electron-hole vertex 
accounting for the exciton condensation in the ladder approximation, 
supplemented by self-energy corrections to electron and hole states. This 
dressing of the bare quasiparticles will have the result of increasing the 
critical coupling at which the excitonic instability takes place, going in the 
same direction as the effect of screening the Coulomb interaction.
We will see however that, incorporating the dynamical polarization in the 
ladder approximation, the screening effects are softened at $N = 4$, 
leading to values of the critical coupling below those corresponding 
to the nominal interaction strength in isolated free-standing 
graphene.

We consider the field theory for Dirac quasiparticles in graphene
interacting through the long-range Coulomb potential, with a Hamiltonian
given by 
\begin{eqnarray}
\lefteqn{H = i v_F \int d^2 r \; \overline{\psi}_i({\bf r}) 
 \mbox{\boldmath $\gamma   \cdot \nabla $} \psi_i ({\bf r}) }   \nonumber \\
  &   &     + \frac{e^2}{8 \pi} \int d^2 r_1
\int d^2 r_2 \; \rho ({\bf r}_1) 
       \frac{1}{|{\bf r}_1 - {\bf r}_2|} \rho ({\bf r}_2)  \;\;\;\;\;
\label{ham}
\end{eqnarray}
where  $\{ \psi_i \}$ is a collection of $N/2$ four-component Dirac 
spinors, $\overline{\psi}_i = \psi_i^{\dagger} \gamma_0 $, and 
$\rho ({\bf r}) = \overline{\psi}_i ({\bf r}) \gamma_0 \psi_i ({\bf r})$.
The matrices $\gamma_{\sigma } $ satisfy 
$\{ \gamma_\mu, \gamma_\nu \} = 2 \: {\rm diag } (1,-1,-1)$
and can be conveniently represented in terms of Pauli matrices as
$\gamma_{0,1,2} = (\sigma_3, \sigma_3 \sigma_1, \sigma_3 \sigma_2) \otimes
 \sigma_3$, where the first factor acts on the two sublattice components of 
the graphene lattice. 

\begin{figure}
\begin{center}
\mbox{\epsfxsize 8.0cm \epsfbox{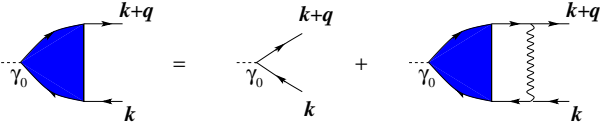}}
\end{center}
\caption{Self-consistent diagrammatic equation for the vertex of
$\langle \rho_m ({\bf q}) \psi ({\bf k}+{\bf q}) \psi^{\dagger} ({\bf k}) 
\rangle$, equivalent to the sum of ladder diagrams built from
the iteration of the Coulomb interaction (wavy line) between electron and
hole states (arrow lines).}
\label{one}
\end{figure}

Our main interest is to study the behavior of the vertex for the operator
$\rho_m ({\bf r}) =  \overline{\psi}({\bf r}) \psi ({\bf r})$. This gives the 
order parameter for the exciton condensation, and the signal that it gets a 
nonvanishing expectation value can be obtained from the divergence  
of $\langle T \rho_m ({\bf q}, t) \rho_m (-{\bf q}, 0) \rangle$.
The singular behavior of this susceptibility can be traced back to the 
divergence of the vertex for 
$\langle \rho_m ({\bf q}) \psi ({\bf k}+{\bf q}) \psi^{\dagger} ({\bf k}) 
\rangle$ at ${\bf q} \rightarrow 0$. We will denote the vertex in this limit 
(setting also the corresponding frequency $\omega_q = 0$) 
by $\Gamma ({\bf k},\omega_k)$. In 
the ladder approximation, depicted diagrammatically in Fig. \ref{one}, the 
vertex is bound to satisfy the equation
\begin{equation}
\Gamma ({\bf k},\omega_k) = \gamma_0 +  
    \int \frac{d^2 p}{(2\pi )^2} \frac{d\omega_p}{2\pi } 
       \frac{\Gamma ({\bf p},\omega_p)}{v_F^2{\bf p}^2+\omega_p^2} 
               V({\bf k}-{\bf p},i\omega_k - i\omega_p)
\label{self}
\end{equation}
where $V({\bf p},\omega_p)$ stands for the Coulomb interaction. We will deal
in general with the RPA to screen the potential, so that 
$V({\bf p}, \omega_p) = e^2/(2 |{\bf p}| + e^2  \chi ({\bf p}, \omega_p))$,
in terms of the polarization $\chi $ for $N$ two-component Dirac fermions.

Eq. (\ref{self}) is formally invariant under a 
dilatation of frequencies and momenta, which shows that the scale
of $\Gamma ({\bf k},\omega_k)$ is dictated by the high-energy cutoff 
$\Lambda $ needed to regularize the integrals. The vertex acquires in general 
an anomalous dimension $\gamma_{\psi^2}$, which governs the behavior under 
changes in the cutoff\cite{amit} 
\begin{equation}
\Gamma ({\bf k},\omega_k) \sim \Lambda^{\gamma_{\psi^2}}
\end{equation}
We recall below how to compute $\gamma_{\psi^2}$, showing that it diverges 
at a critical value of the interaction strength. This 
translates into a divergence of the own susceptibility at momentum
transfer ${\bf q} \rightarrow 0$, providing then the signature
of the condensation of $\rho_m ({\bf r}) =  \overline{\psi}({\bf r}) \psi ({\bf r})$
and the consequent development of the excitonic gap.

{\em Self-energy corrections to ladder approximation.---}
We deal first with the approach in which electrons and holes are dressed by 
self-energy corrections, while the Coulomb interaction in (\ref{self}) is 
screened by means of the static RPA with 
$\chi ({\bf p}, 0) = (N/16)|{\bf p}|/v_F $.
It is known that graphene remains a 
conventional Fermi liquid even at the charge neutrality point, with a 
quasiparticle weight that does not vanish at the Fermi level\cite{prbr}. 
The most important self-energy effect comes from the renormalization of the 
Fermi velocity at low energies\cite{np2}, and this is the feature that we want 
to incorporate in our analysis, identifying $v_F$ in Eq. (\ref{self}) with 
the Fermi velocity dressed by self-energy corrections.

The electron self-energy corrections, as well as the terms of the 
ladder series, are given by logarithmically divergent integrals that need to
be cut off at a given scale $\Lambda $. Alternatively, one can also define
the theory at spatial dimension $d = 2 - \epsilon$, what automatically 
regularizes all the momentum integrals. Eq. (\ref{self}) then becomes
\begin{equation}
\Gamma ({\bf k},0) = \gamma_0 +  \frac{e_0^2}{4\kappa } 
   \int \frac{d^d p}{(2\pi )^d} \Gamma ({\bf p},0) 
    \frac{1}{\widetilde{v}_F({\bf p}) |{\bf p}|} \frac{1}{|{\bf p}-{\bf k}|}
\label{selfcons}
\end{equation}
where $\widetilde{v}_F ({\bf p})$ is the Fermi velocity dressed with the 
self-energy corrections, $\kappa = 1 + N e^2/32 v_F$, and $e_0^2$ is related 
to $e^2$ through an auxiliary momentum scale $\rho $ such that 
$e_0^2 = \rho^{\epsilon} e^2 $. 

In the ladder approximation, the Fermi velocity gets a divergent correction
only from the ^^ ^^ rainbow'' self-energy diagram with exchange of a single 
screened interaction\cite{np2}. The dressed Fermi velocity 
becomes
\begin{equation}
\widetilde{v}_F({\bf p}) = v_F + \frac{e_0^2}{\kappa} 
 \frac{1}{(4\pi )^{2 - \epsilon /2}}   
  \frac{\Gamma (\tfrac{\epsilon }{2}) \Gamma (\tfrac{1-\epsilon}{2}) 
                                            \Gamma (\tfrac{3-\epsilon}{2}) }
  {  \Gamma (2 - \epsilon) }
   \frac{1}{|{\bf p}|^{\epsilon}}
\label{vd}
\end{equation}
The expressions (\ref{selfcons}) and (\ref{vd}) are singular in the limit 
$\epsilon \rightarrow 0$. The most convenient way to show that all 
the divergences can be renormalized away is to resort at this point to a 
perturbative computation of $\Gamma ({\bf k},0)$.

The solution of (\ref{selfcons}) can be obtained in the form
\begin{equation}
\Gamma ({\bf k},0) = 
 \gamma_0 
 \left(1 + \sum_{n=1}^{\infty} \lambda_0^n 
                           \frac{r_n }{|{\bf k}|^{n\epsilon}} \right)
\end{equation}
with $\lambda_0 = e_0^2/4\pi \kappa v_F $.
Each term in the sum can be obtained from the previous one by noticing that
\begin{eqnarray}
\lefteqn{ \int \frac{d^d p}{(2\pi )^d} \frac{1}{|{\bf p}|^{(m-1)\epsilon} }
            \frac{1}{|{\bf p}|} \frac{1}{|{\bf p}-{\bf k}|}            }
                                                            \nonumber   \\
  & \;\;\;\;\;\;\;\;\;\;\;\;\;\;\;\;\;\; 
 =  \frac{(4\pi )^{\epsilon /2}}{4 \pi^{3/2}}   
  \frac{\Gamma (\frac{m\epsilon }{2}) \Gamma (\frac{1-m\epsilon}{2}) 
                                            \Gamma (\frac{1-\epsilon}{2}) }
  { \Gamma (\frac{1+(m-1)\epsilon}{2}) \Gamma (1-\frac{m + 1}{2}\epsilon) }
 \frac{1}{|{\bf k}|^{m\epsilon}}   &
\end{eqnarray}
At each given perturbative level, the vertex displays then a number of poles in 
the variable $\epsilon $. The crucial point is that these divergences can be
reabsorbed by passing to physical quantities such that 
$v_F = Z_v(v_F)_{\rm ren}$ and 
$\overline{\psi} \psi = Z_{\psi^2} (\overline{\psi} \psi )_{\rm ren}$
(the scale of the single Dirac field is not renormalized in the ladder 
approximation).

The renormalized vertex
$\Gamma_{\rm ren} = Z_{\psi^2} \Gamma $ can be actually made finite with a 
suitable choice of momentum-independent factors $Z_v$ and $Z_{\psi^2}$. 
$Z_v$ must be chosen to cancel the $1/\epsilon $ pole in (\ref{vd}), and it 
has therefore the simple structure $Z_v = 1 + b_1 /\epsilon  $, with 
$b_1 = - e^2/16\pi \kappa (v_F)_{\rm ren}$. On the other hand, we have the 
general structure
\begin{equation}
Z_{\psi^2} = 1 + \sum_{n=1}^{\infty} \frac{c_n }{\epsilon^n}
\label{poles}
\end{equation}
The position of the different poles must be chosen to enforce the
finiteness of $\Gamma_{\rm ren} = Z_{\psi^2} \Gamma $ in the limit
$\epsilon \rightarrow 0$. The computation of the 
first orders of the expansion gives for instance the result
\begin{eqnarray}
c_1 (\lambda ) & = &  - \frac{1}{2} \lambda - \tfrac{1}{8} \log(2) \: \lambda^2 
      - \tfrac{1}{1152} \left( \pi ^2 + 120 \log ^2(2) \right)   \lambda^3     \nonumber   \\
  &  &   -  \tfrac{10 \pi ^2 \log (2)+688 \log ^3(2)+15 \zeta (3)}{6144}   \lambda^4  
                                                               \nonumber  \\
 & & - \tfrac{13 \pi ^4+2064 \pi ^2 \log ^2(2)+144 \left(716 \log ^4(2)+37 \log (2) \zeta (3)\right)}{737280} \: \lambda^5    
                                                                             \nonumber \\
  &  &                                                +    \ldots             \nonumber \\
c_2 (\lambda ) & = &  \tfrac{1}{16} \: \lambda^2 + 
 \tfrac{1}{24} \log(2) \: \lambda^3 + \tfrac{1}{18432} \left( 5 \pi ^2 + 744 \log ^2(2) \right)  \lambda^4  \nonumber  \\
   &  &   + \tfrac{110 \pi ^2 \log (2)+8592 \log ^3(2)+135 \zeta (3)}{184320}  \: \lambda^5 
                                                         +  \ldots         \nonumber  \\
c_3 (\lambda ) & = & - \tfrac{1}{768} \log (2) \: \lambda^4 - \tfrac{1}{184320} \left( \pi ^2+360 \log ^2(2) \right)  \lambda^5 
                                                             + \ldots   \nonumber  \\
c_4 (\lambda ) & = & - \tfrac{1}{7680} \log (2) \: \lambda^5  + \ldots  
\label{coeff}
\end{eqnarray}
where the series are written in terms of the renormalized coupling
$\lambda \equiv \rho^{-\epsilon} Z_v \lambda_0$

The physical observable in which we are interested is the anomalous 
dimension $\gamma_{\psi^2}$. The change in the dimension of $\Gamma_{\rm ren}$
comes from the dependence of $Z_{\psi^2}$ on the only dimensionful scale 
$\rho $, being 
$\gamma_{\psi^2} = (\rho /Z_{\psi^2}) \partial Z_{\psi^2} /\partial \rho $ \cite{amit}. 
The bare theory at $d \neq 2$ does not know about the arbitrary scale 
$\rho $, and the 
independence of $\lambda_0 = \rho^{\epsilon} \lambda /Z_v $ on that
variable leads to 
\begin{equation}
\rho \frac{\partial \lambda }{\partial \rho } = 
 - \epsilon \lambda - \lambda b_1 (\lambda )
\label{rge}
\end{equation}
The anomalous dimension is then\cite{ram}
\begin{equation}
\gamma_{\psi^2} = \rho  
  \frac{\partial \lambda }{\partial \rho }
  \frac{\partial \log Z_{\psi^2} }{\partial \lambda }
 = - \lambda c_1' (\lambda )
\label{dreg}
\end{equation}
In the derivation of (\ref{dreg}), it has been implicitly assumed that poles 
in the variable $\epsilon $ cannot appear at the right-hand-side of the 
equation. For this to be true, the set of equations 
$c_{n+1}' = c_n c_1' - b_1 c_n' $ must be satisfied identically\cite{ram}. 
Quite remarkably, we have verified that this is the case, up to the order 
$\lambda^7 $ we have computed the coefficients in (\ref{poles}). This is the proof of
the renormalizability of the theory, which guarantees that physical quantities
like $\gamma_{\psi^2}$ remain finite in the limit $\epsilon \rightarrow 0$.

\begin{figure}[t]
\begin{center}
\mbox{\epsfxsize 5.0cm \epsfbox{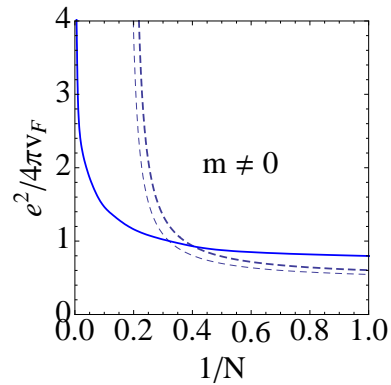}}
\end{center}
\caption{Phase diagram showing the boundary of the phase with exciton
condensation ($m \neq 0$), obtained in the ladder approximation with and 
without electron self-energy corrections (upper and lower dashed lines) and 
with dynamical screening of the Coulomb interaction (full line).
}
\label{two}
\end{figure}

From the practical point of view, the important result is the evidence that 
the perturbative expansion of $c_1 (\lambda )$ approaches a geometric series 
in the $\lambda $ variable. It can be checked that the coefficients in the 
expansion grow exponentially with the order $n$, in such a way that
\begin{equation}
- c_1 (\lambda ) \geq \sum_{n=1}^{\infty} d^n \lambda^n 
\end{equation}
A lower bound for $d$ can be obtained from the first orders in
$c_1 (\lambda )$. This estimate tends to increase as it is made
from higher orders in the expansion. 
The assumption of 
scaling with the order $n$ allows us to estimate a radius
of convergence $\lambda_c \approx 0.49$ (to be compared with the value in the 
approach neglecting self-energy corrections, which leads to 
$\lambda_c \approx 0.45$ \cite{me}, in close agreement with the result of 
Ref. \onlinecite{gama}). The critical coupling in the variable 
$\lambda = \alpha /\kappa $ can be used to draw the boundary for exciton 
condensation in the $(N, \alpha )$ phase diagram, represented in Fig. \ref{two}.
For $N = 4$, we get in particular the critical coupling $\alpha_c \approx 2.09$,
significantly above the critical value that would be obtained from the radius
of convergence without self-energy corrections ($\alpha_c \approx 1.53$).

{\em Dynamical screening in the ladder approximation.---}
In the framework of the ladder approximation, one can also study the effect
of the dynamical screening of the Coulomb interaction. We can go beyond
the static RPA by taking into account the full effect of the 
frequency-dependent polarization, which for Dirac fermions 
takes the form
$\chi ({\bf p}, \omega_p) = 
(N/16) {\bf p}^2/\sqrt{v_F^2 {\bf p}^2 - \omega_p^2}$ \cite{np2}.
This expression can be introduced in Eq. (\ref{self}) to look again for
self-consistent solutions for the vertex $\Gamma ({\bf k},\omega_k)$. 
While this problem
does not afford an analytic approach of the type shown before, one can
resort to numerical methods for the resolution of the integral
equation. In this procedure, we find again that there is a critical 
coupling at which $\Gamma ({\bf k},\omega_k)$ blows up, marking the boundary 
between two different regimes where the integral equation has respectively
positive and negative (unphysical) solutions.

In practice, we have solved the integral equation (\ref{self}) by defining the
vertex in a discrete set of points in frequency and momentum space. One can 
take as independent variables in $\Gamma ({\bf k},\omega_k)$ the modulus of 
${\bf k}$ and positive frequencies $\omega_k $, and we have adopted accordingly 
a grid of dimension $l \times l$ covering those variables. The advantage of 
this approach is that the number $l$ plays the role of cutoff, making 
straightforward to assess the effect of higher frequencies and momenta as $l$ 
is increased. We have solved the integral equation in $l \times l$ grids with 
$l$ running up to a value of 200, for which it is still viable to invert a 
matrix of dimension $l^2$. One can rely moreover on 
the scale invariance of the theory to find the trend at larger values of $l$,
as the critical coupling $\alpha_c$ must obey a well-defined finite-size scaling 
law as a function of the 
cutoff  $\alpha_c (l) = \alpha_c (\infty ) + c/l^{\nu }$.

For a given value of $N$, we have determined the critical point at which the 
vertex $\Gamma ({\bf k},\omega_k)$ blows up. The value of $\alpha_c (l)$ 
for our largest $l$ provides an upper bound for the critical
coupling in the continuum limit, as $\alpha_c (l)$ turns out to be always a
decreasing function of the variable $l$. On the other hand, the use of the 
above finite-size scaling law allows to estimate $\alpha_c (\infty )$ .
We have chosen to represent in Fig. \ref{two}
the conservative upper bound $\alpha_c (200)$ as a function 
of $N$. In marked difference with other approaches, we observe that now a 
critical coupling always exists, no matter how large the value of $N$ may be. 
For $N = 4$ corresponding to graphene, we get the values 
$\alpha_c (200) \approx 1.08$ and $\alpha_c (\infty ) \approx 0.99$.

We see that the value of the critical coupling obtained upon dynamical 
screening of the interaction is substantially smaller than the 
value found for $N = 4$ in the static approximation. This agrees with the 
results obtained in Ref. \cite{ggg}, where the resolution of the gap equation
was accomplished taking into account the frequency-dependent polarization. 
The critical coupling obtained there for $N = 4$, $\alpha_c \approx 0.92$,
is actually very close to our estimate $\alpha_c (\infty ) \approx  0.99$.
These values are also close to the critical coupling 
$\alpha_c \approx 1.08$ found in lattice Monte Carlo 
simulations\cite{drut1}, suggesting that the consideration of the 
dynamical screening provides a most sensible approximation to the excitonic 
instability.

{\em Conclusion.---}
We have shown that the various many-body corrections used to dress the 
electron quasiparticles and 
the Coulomb interaction can have significant impact on the excitonic
instability. The effects of the electron self-energy corrections and the
electron-hole polarization have an important role in reducing the strength of
the instability. We have seen however
that, as anticipated in Ref. \cite{ggg}, the simple static approximation 
overestimates the screening effects, and that a more accurate approach to the
problem requires the consideration of the dynamical screening of the 
interaction.

It is puzzling that, if we were to use the nominal values of the parameters 
appropriate for graphene, we would arrive at the conclusion that an isolated 
free-standing layer of the material (for which $\alpha \approx 2.2$) should be
in the phase of exciton condensation. This is at odds with the absence of any 
experimental observation of a gap in suspended graphene samples. Our many-body 
analysis shows that the only possible relevant effects that have been 
dismissed are those related to the scaling of the quasiparticle parameters. 
In this respect, the growth of the Fermi velocity at low 
energies\cite{prbr} can have a deep impact to prevent the excitonic 
instability\cite{sabio,see}. This effect, expressed by the scaling law (\ref{rge})
at $\epsilon = 0$, has been already observed in experimental samples of graphene
at very low doping levels\cite{paco}. It is quite plausible that, at the 
low-energy scales where the gap could develop (about three orders of magnitude
below the scale of the high-energy cutoff), the scaling of the Fermi
velocity may have driven the effective coupling to such small values that 
the excitonic instability cannot then proceed. This can be one more consequence of 
the nontrivial scaling properties of the theory of Dirac fermions, 
implying that the electrons in graphene approach a noninteracting
regime as they are observed at energies arbitrarily closer to the charge 
neutrality point.

{\em Acknowledgments.---}
We thank F. Guinea for very useful discussions.
The financial support from MICINN (Spain) through grant
FIS2008-00124/FIS is also acknowledged.


\begin{thebibliography}{99}




\bibitem{geim}
K. S. Novoselov, A. K. Geim, S. V. Morozov, D. Jiang, M. I. Katsnelson,
I. V. Grigorieva, S. V. Dubonos and A. A. Firsov, Nature {\bf 438}, 197 (2005).

\bibitem{kim}
Y. Zhang, Y.-W. Tan, H. L. Stormer and P. Kim, Nature {\bf 438}, 201 (2005).

\bibitem{rmp}
A. H. Castro Neto, F. Guinea, N. M. R. Peres, K. S. Novoselov and
A. K. Geim, Rev. Mod. Phys. {\bf 81}, 109 (2009).

\bibitem{nil}
V. M. Pereira, J. Nilsson and A. H. Castro Neto, Phys. Rev. Lett. {\bf 99}, 
166802 (2007).

\bibitem{fog}
M. M. Fogler, D. S. Novikov, and B. I. Shklovskii, Phys. Rev. B {\bf 76}, 
233402 (2007).

\bibitem{shy}
A. V. Shytov, M. I. Katsnelson, and L. S. Levitov, 
Phys. Rev. Lett. {\bf 99}, 236801 (2007).


\bibitem{ter}
I. S. Terekhov, A. I. Milstein, V. N. Kotov, and O. P. Sushkov,
Phys. Rev. Lett. {\bf 100}, 076803 (2008).
 

\bibitem{appel}
T. Appelquist, D. Nash and L. C. R. Wijewardhana, Phys. Rev. Lett. {\bf 60},
2575 (1988).

\bibitem{khves}
D. V. Khveshchenko, Phys. Rev. Lett. {\bf 87}, 246802 (2001).

\bibitem{gus}
E. V. Gorbar, V. P. Gusynin, V. A. Miransky and I. A. Shovkovy, Phys. Rev. 
B {\bf 66}, 045108 (2002).

\bibitem{ale}
I. L. Aleiner, D. E. Kharzeev and A. M. Tsvelik, Phys. Rev. B {\bf 76},
195415 (2007).

\bibitem{son}
J. E. Drut and D. T. Son, Phys. Rev. B {\bf 77}, 075115 (2008).

\bibitem{vafek}
O. Vafek and M. J. Case, Phys. Rev. B {\bf 77}, 033410 (2008).  


\bibitem{drut1}
J. E. Drut and T. A. L\"ahde, Phys. Rev. Lett. {\bf 102}, 026802 (2009);
Phys. Rev. B {\bf 79}, 241405(R) (2009).

\bibitem{hands}
See also S. J. Hands and C. G. Strouthos, Phys. Rev. B {\bf 78}, 165423 (2008);
W. Armour, S. Hands, C. Strouthos, Phys. Rev. B {\bf 81}, 125105 (2010).

\bibitem{gama}
O. V. Gamayun, E. V. Gorbar and V. P. Gusynin, Phys. Rev. B {\bf 80},
165429 (2009).

\bibitem{fer}
J. Wang, H. A. Fertig and G. Murthy, Phys. Rev. Lett. {\bf 104}, 186401 (2010).

\bibitem{me}
J. Gonz\'alez, Phys. Rev. B {\bf 82}, 155404 (2010).

\bibitem{brey}
J. Wang, H. A. Fertig, G. Murthy and L. Brey, Phys. Rev. B {\bf 83}, 
035404 (2011).

\bibitem{ggg}
O. V. Gamayun, E. V. Gorbar and V. P. Gusynin, Phys. Rev. B {\bf 81},
075429 (2010).

\bibitem{khves2}
D. V. Khveshchenko, J. Phys.: Condens. Matter {\bf 21}, 075303 (2009).

\bibitem{sabio}
J. Sabio, F. Sols and F. Guinea, Phys. Rev. B {\bf 82}, 121413͑(R) (2010).


\bibitem{amit}
D. J. Amit and V. Mart\'{\i}n-Mayor, {\em Field Theory, the Renormalization 
Group, and Critical Phenomena}, Chaps. 6 and 8 
(World Scientific, Singapore, 2005).

\bibitem{prbr}
J. Gonz\'alez, F. Guinea and M. A. H. Vozmediano,
Phys. Rev. B {\bf 59}, R2474 (1999).

\bibitem{np2}
J. Gonz\'alez, F. Guinea and M. A. H. Vozmediano,
Nucl. Phys. B {\bf 424}, 595 (1994).

\bibitem{ram}
P. Ramond, {\em Field Theory: A Modern Primer}, Chap. IV (Benjamin/Cummings, Reading, 1981).


\bibitem{see}
See also I. F. Herbut, V. Juri\v{c}i\'c and O. Vafek, Phys. Rev. B {\bf 80}, 075432 (2009); 
V. Juri\v{c}i\'c, I. F. Herbut and G. W. Semenoff, Phys. Rev. B {\bf 80}, 081405 (2009).

\bibitem{paco}
D. C. Elias, A. S. Mayorov, F. Guinea, K. S. Novoselov and A. K. Geim, in preparation.








\end{thebibliography}
\end{document}